\documentclass[journal=jpccck,manuscript=article]{achemso}
\usepackage{amsmath,color}
\usepackage{graphicx}
\usepackage{amssymb}

\author{Sandeep K. Jain}
\affiliation[Institute for Theoretical Physics,Universiteit Utrecht]
{Institute for Theoretical Physics, Universiteit Utrecht,
Princetonplein 5, 3584 CC Utrecht, The Netherlands}
\email{sandeepiitr7@gmail.com}

\author{Gerard T. Barkema}
\affiliation[Department of Information and Computing Science, Universiteit Utrecht]
{Department of Information and Computing Science, Universiteit Utrecht,
Princetonplein 5, 3584 CC Utrecht, The Netherlands}
\email{G.T.Barkema@uu.nl}

\title{Rupture of amorphous graphene via void formation}

\begin{document}


\begin{abstract}
Apart from its unique and exciting electronic properties, many sensor based applications of graphene are purely based on its mechanical and structural properties. Here we report a numerical and analytical study of a void in amorphous (small domain polycrystalline) graphene, and show that the energetics of a void is a balance between the line tension cost versus the increased area gain. Using the concepts of classical nucleation theory, we show that the critical radius of a void formed in amorphous graphene at constant pressure is simply the ratio of line tension at the void and the applied pressure. The values of the critical radius of the void for flat and buckled graphene are 3.48\AA~and 3.31\AA, respectively at 2 eV/\AA${^2}$ pressure. We also show that the dominant finite size correction to the line tension is inversely proportional to the radius of the void in both flat and buckled cases. Contrary to conventional wisdom, with the help of a simple analytical model we find that the shear modulus sets the lower limit of the line tension in the samples. This makes our study relevant for other two-dimensional amorphous materials such as h-BN, phosphorene, borophene, and transition metal dichalcogenides. Our results are useful for the better understanding of polycrystalline graphene under tension and therefore have direct implications on the very fascinating field of strain engineering known as "straintronics" to manipulate or improve graphene's properties.
\end{abstract}

\section{INTRODUCTION}

The discovery of graphene has provoked a revolution in nanotechnology, as the structural, thermal, and electronic properties of graphene make it a very useful component for a large variety of devices \cite{Neto2009,Geim2009,Geim2007,Dolleman2016}. Most of these applications require a high quality, large graphene samples. The quality of the samples is very important for the observation of the special features of graphene, such as ultrahigh electron mobility \cite{Bolotin2008,Morozov2008,Mayorov2011}, very high thermal conductivity \cite{Balandin2011}, half integer quantum-Hall effect \cite{Zhang2005,Novoselov2007}, massless Dirac fermions \cite{Novoselov2005} as well as for its mechanical and chemical properties, e.g., its permeability \cite{Nair2012} and very high Young's modulus \cite{Lee2008,Booth2008}. Graphene is claimed to be the strongest material ever, as it has extraordinary elastic properties. 

Recently, with the help of the chemical vapor deposition (CVD) technique; synthesis and production of large graphene samples became possible \cite{Zhang2013}. However, in reality, chemical vapor deposition (CVD) grown graphene is a polycrystalline material, hence contains many defects, and naturally buckles out of the crystalline plane \cite{Yazyev2014,Aissa2015,Lee2015,Deng2016}. These defects limit the visibility of the special properties of pristine graphene, but on the other hand have positive effects when graphene is used as an anode material for metal-ion batteries \cite{Yu2013}. Study of lattice defects in graphene is of both fundamental and practical relevance, since they are inevitably present in the samples and have direct consequences on the physical and chemical properties, thereby impeding its applications in the semiconductor industry \cite{Banhart2011}. Amorphous two-dimensional carbon structures are also very important since the ordered hexagon arrangement of carbon atoms in graphene is not directly suitable for many practical studies and applications like chemical sensor and nanoelectronics \cite{Kotakoski2015}. 

Since graphene is a one-atom thick two-dimensional layer, deformations via stretching offer very tempting aspects to control or manipulate its properties \cite{Pereira2009a,Pereira2009,Ni2008,Kim2009,Bao2009}. Graphene is known to be very flexible and can be stretched by as much as 20\% without inducing defects and rupture \cite{Si2016}. It has been shown that in graphene, strain can induce many fascinating effects such as pseudo-magnetic fields greater than 300 T at room temperature \cite{Levy2010}, zero field quantum Hall effect \cite{Guinea2010}, enhancement of electron-phonon coupling \cite{Si2013}, and superconducting states in pseudo-Landau-levels \cite{Uchoa2013}. Defects combined with strain is also very interesting, as they lead to some new effects, not present otherwise \cite{Lu2013,Lu2015}. Mechanical properties of large domain polycrystalline graphene samples have been reported in literature \cite{Kotakoski2012,Song2013,Mortazavi2014,Sha2014,Sha2014a,Shekhawat2016}. Various empirical methods were used to study the deformations and tensile properties in graphene, carbon nanotubes and the amorphous carbon structures \cite{Jensen2015,Lindsay2010,Hossain2018,Zhao2009,Memarian2015}. A comprehensive and detailed study of the effect of stretching on the structural properties of amorphous graphene is missing.

Very recently, our preliminary results showed that mechanical stretching also has an interesting and significant impact on the polycrystalline properties of the graphene. We observed that via regulated stretching, one can suppress the point and line defects in the samples and hence ultimately can have a single crystalline domain in graphene. An excess amount of stretching, however can rupture the sample via void formation. It is, therefore, of both practical and fundamental importance to address the effect of stretching on the structural properties of the amorphous graphene. Experimentally also it is very relevant since during the synthesis, production, and transfer process structural strains are expected to be present in the samples due to the surface corrugations of the substrate or the lattice mismatch between substrate and graphene layer. In this article, we report the energetics of void formation in amorphous graphene.

\section{METHOD}

In our simulations, we use a recently-developed semi-empirical potential for graphene \cite{Jain2015} given by 
\begin{equation} \label{eq:Potential}
E=\frac{3}{16}\frac{\alpha}{d^2}\sum_{i,j}(r^2_{ij}-d^2)^2+\frac{3}{8}\beta d^2 \sum_{j,i,k}(\theta_{jik}-\frac{2 \pi}{3})^2 + \gamma \sum_{i,jkl}r^2_{i,jkl}.
\end{equation}
Here, $r_{ij}$ is the length of the bond between two atoms $i$ and $j$, $\theta_{jik}$ the angle between the two bonds connecting atom $i$ to $j$ and $k$, respectively. $r_{i,jkl}$ is the distance between atom $i$ and the plane through the three neighboring atoms $j$, $k$ and $l$ connected to atom $i$. The parameters $\alpha=26.060\text{ eV/\AA}^2$, $\beta=5.511\text{ eV/\AA}^2$, $\gamma=0.517\text{ eV/\AA}^2$, and $d=1.420\text{ \AA}$ are used, which were obtained from density-functional theory (DFT) calculations \cite{Jain2015}. This potential was effectively used to study vibrational properties of graphene \cite{Jain2015a} and carbon nanotubes \cite{Pool2017}, effect of boundary conditions in graphene samples \cite{Jain2016}, structure of twisted bilayer graphene \cite{Jain2016a}, and graphene nanobubbles \cite{Jain2017}.

\section{RESULTS AND DISCUSSION}

To generate an unbiased isotropic three-fold connected random network, we use an approach based on  Voronoi networks \cite{Voronoi1908}. We place N/2 random points in a periodic box, and determine the Voronoi cells around each of these random points. The boundaries between two neighboring cells are then transformed into carbon-carbon bonds, and at the locations where three Voronoi cells meet, we place a carbon atom. Note that the initial random points cease to play a role. The resulting highly strained network with N carbon atoms is then relaxed. In our simulations we use force free (FF) boundary conditions \cite{Jain2016}. \\

\begin{figure*}
\centering
\includegraphics[width=1.0\textwidth]{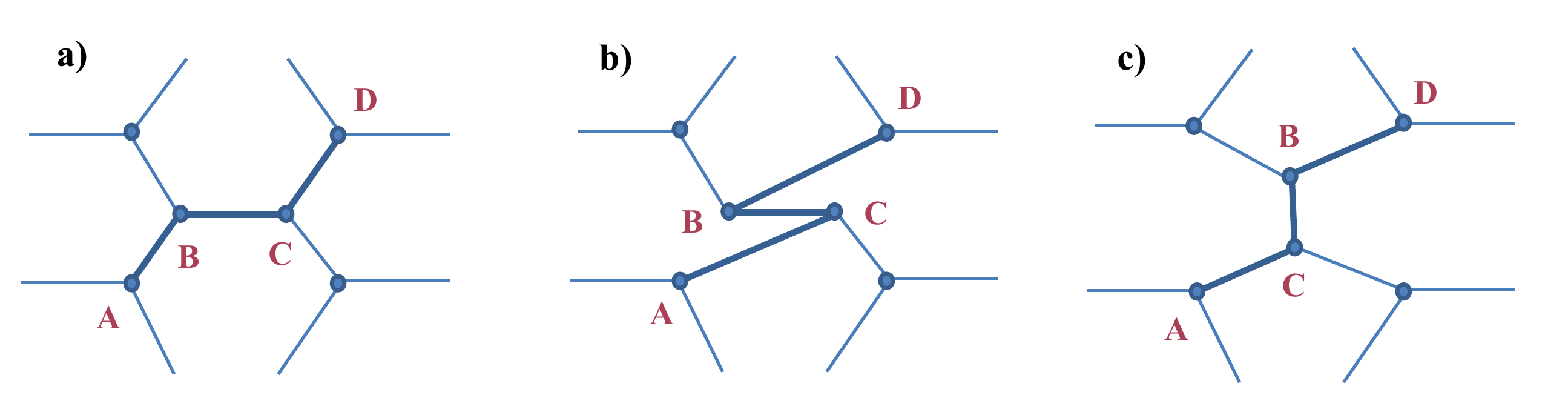}
\caption{Bond transposition mechanism in crystalline graphene. a) A string of four random atoms are selected in the sample to switch the bonds (A--B and C--D). b) In a move a connection is made between atoms A and C and similarly between atoms B and D, replacing bonds between atoms A and B as well as C and D. c) The sample is then relaxed. }
\label{Fig-1}
\end{figure*}

Rupture is a dynamical phenomenon, hence we need to specify our dynamics. Per unit of time, we propose N bond transpositions, i.e. there is a fixed rate at which each possible bond switch is attempted. In our simulations, we start with an initial three-fold coordinated amorphous graphene sample with $5000$ carbon atoms, generated with the Voronoi approach described above. As these networks are highly strained, we first do a quick relaxation over 5 units of time via Monte Carlo dynamics with bond transposition moves as illustrated in Figure  \ref{Fig-1}. From there, we study the time evolution under further Monte Carlo dynamics. In more detail, we use the improved bond-switching algorithm in our simulations \cite{Barkema2000}, using our empirical potential (Eq: \ref{eq:Potential}) to describe its energy. A random switch is made in the explicit list of bonds, after which the coordinates are adjusted to the minimal energy state. The new configuration is accepted with a Metropolis probability given by
\begin{equation} \label{eq:Metropolis}
P=\min\left[1,\exp\left(\frac{E_b-E_f}{k_BT}\right)\right].
\end{equation}
Here $E_{b}$ is the energy of the system before the bond transposition, $E_{f}$ is the energy after the bond transposition and $k_BT$ is the thermal energy of the system (0.083 eV). We apply a continuous stretching pressure ($P$) at the system and the stretching energy of the system is given by the following term and added to the potential (Eq:\ref{eq:Potential}) while relaxing the system
\begin{equation} \label{eq:Stretching}
E_{s}=-P  L_{x}  L_{y}  sin(\theta).
\end{equation}

Here, $P$ is the pressure applied on the sample, $L_{x}$, $L_{y}$ are the periodicity vectors, and $\theta$ is the angle between these periodicity vectors.

\begin{figure*}
\includegraphics[width=1.10\textwidth]{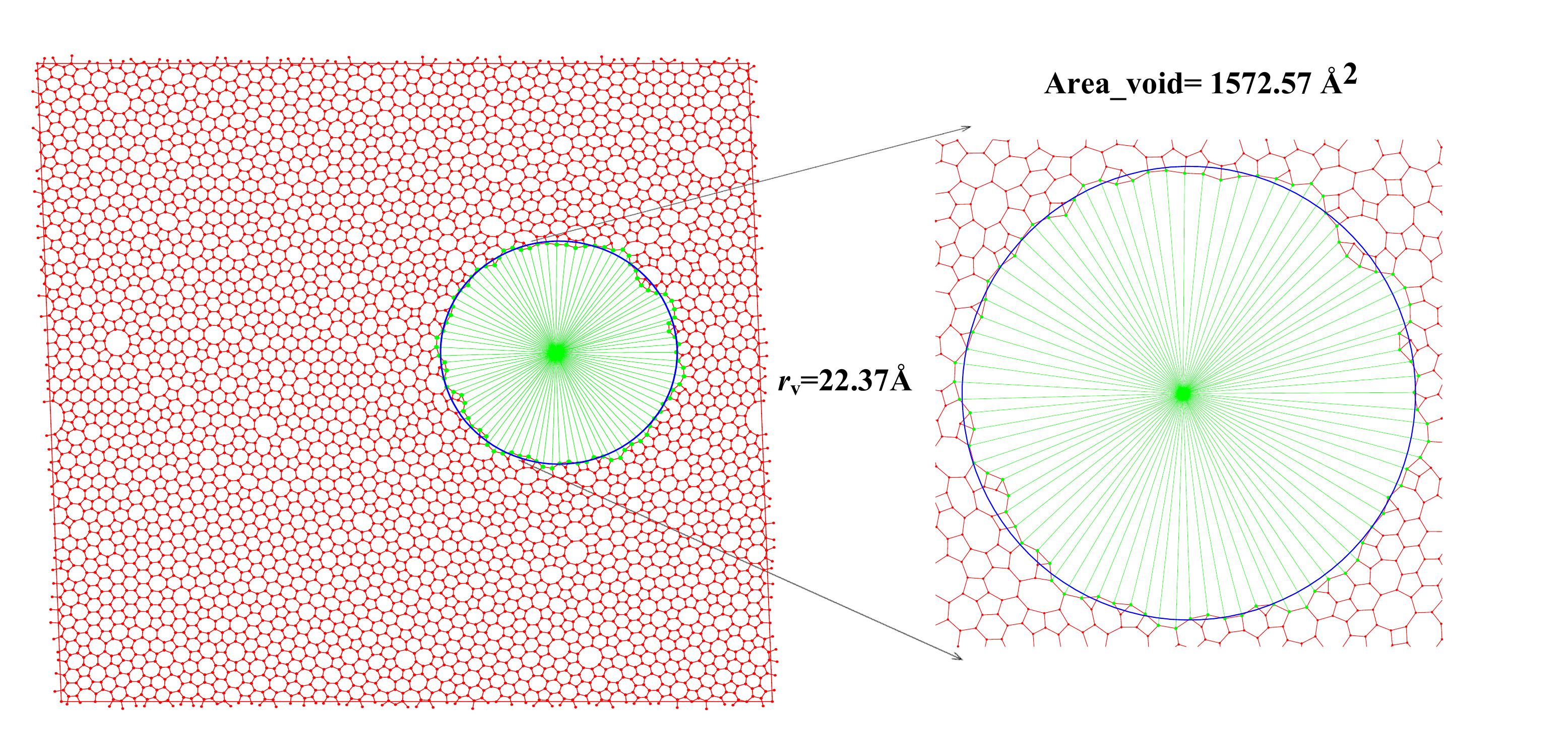}
\caption{Structure of a void in an amorphous graphene sample with $5000$ carbon atoms at a constant pressure of $P=2\text{ eV/\AA}^2$ (Eq. \ref{eq:Stretching}). Here we show an effective method to determine the radius of the void ($r_v$) which we used in our calculations. The whole area of the void ($a_v$) is divided into small triangles connecting to two consecutive atoms at periphery of the void from a point inside the void (as shown in green colored lines). The summation of all the area contribution from these small triangles resulted into the total effective area of the void ($a_v$) and then the effective radius of the void (shown in the blue colored circle in the zoomed in figure) is calculated by $a_v=\pi \times r_{v}^{2}$.}
\label{Fig-2}
\end{figure*}

As the system evolves via bond transposition moves at constant pressure, we observe the formation of void at higher order of the carbon rings ($n \geq 15$) in amorphous graphene. To calculate the effective radius of the void ($r_{v}$), we first calculate the total area inside the void. For that we identify the carbon atoms at the periphery of the void and divide the whole area into small triangles as shown in Figure \ref{Fig-2}. The total area is equated with the area of a circle to extract the effective radius of the void. Notably, we observe lots of three-membered rings around the periphery of the void. Interestingly, if a crystalline sample is stretched well beyond 20\%, our bond transposition dynamics cause it to rupture while retaining three-fold coordination. \\  

\textbf{Energetics of the void}
\\
For a small void, it is energetically unfavorable to grow because of the line tension. Beyond a certain critical size, however, the pressure dominates the line tension, as the first should be multiplied by the boundary length and the second by the area, and the void starts to grow. Using the concepts of classical nucleation theory, we analytically determine the critical radius of the void in a two-dimensional system. At the constant stretching combined with the bond switching, critical radius of the void is defined as the particular size of the void from where the void will only grow in size. 
\\  
The free energy cost due to the line tension ($F_s$) in the sample can be given as
\begin{equation} \label{eq:line_energy}
F_{s}= 2 \pi r_{v}\sigma(r_v)
\end{equation}
where $r_v$ is the radius of the void and $\sigma(r_v)$ is the line tension along the void (free energy per unit length in eV/\AA). 
\\
The free energy gain due to the strain relieved from stretching ($F_a$) can be written as
\begin{equation} \label{eq:strain_energy}
F_{a}= - \pi r_{v}^{2} P
\end{equation}
where $P$ is the pressure acting on the sample in eV/\AA${^2}$.
\\
The total free energy (summation of line energy and strain energy) as a function of the void size ($r_{v}$) can be written as 
\begin{equation} \label{eq:total_energy}
F(r_v)= 2 \pi r_{v}\sigma(r_v) - \pi r_{v}^{2} P.
\end{equation}

\begin{figure*}
\includegraphics[width=0.98\textwidth]{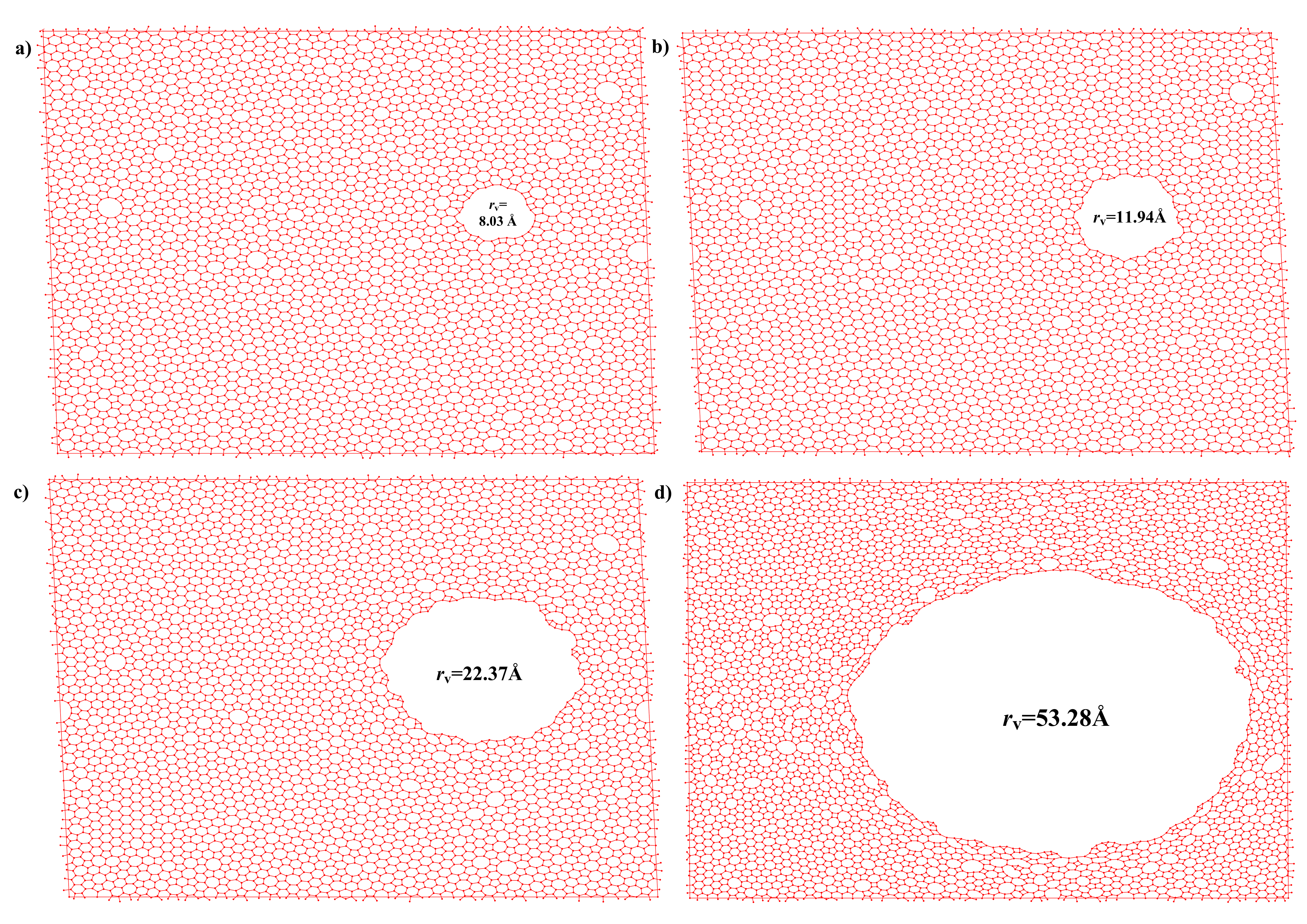}
\caption{Structure and evolution of a void in amorphous graphene, simulated at a constant pressure of $P=2\text{ eV/\AA}^2$. We start with an amorphous sample of graphene and evolve it via bond transposition moves with Metropolis acceptance probability (Eq. \ref{eq:Metropolis}) using the Hamiltonian given in Eq. \ref{eq:Potential}. This method is widely used to study the dynamics and evolution of amorphous materials such as a-Si \cite{Barkema2000}. Here we show various snapshots of the graphene sample captured during the evolution process. The radius of the voids are (a) $r_v$ = 8.03\AA. (b) $r_v$ = 11.94\AA. (c) $r_v$ = 22.37\AA. (d) $r_v$ = 53.28\AA. Once the critical radius of the void is crossed, the size of the void will keep on increasing as shown in the figure, and sample will start to rupture along the void.}
\label{Fig-3}
\end{figure*}

The formation of the void and the evolution of the structure (various snapshots at constant pressure) in $5000$ carbon atom sample is shown in Figure \ref{Fig-3}. The radius of the void ($r_{v}$) continuously increases during the bond transpositions at constant pressure ($P=2 ~\text{ eV/\AA}^2$), once the void crosses the value of the critical radius. We start with a highly amorphous graphene sample with a void ($r_{v}=8.03$ ~\AA) as shown in Figure \ref{Fig-3} a). As the pressure is continuously applied on the sample together with bond transposition, the whole sample starts to rupture along the void. 

\begin{figure*}
\centering
\includegraphics[scale=0.46]{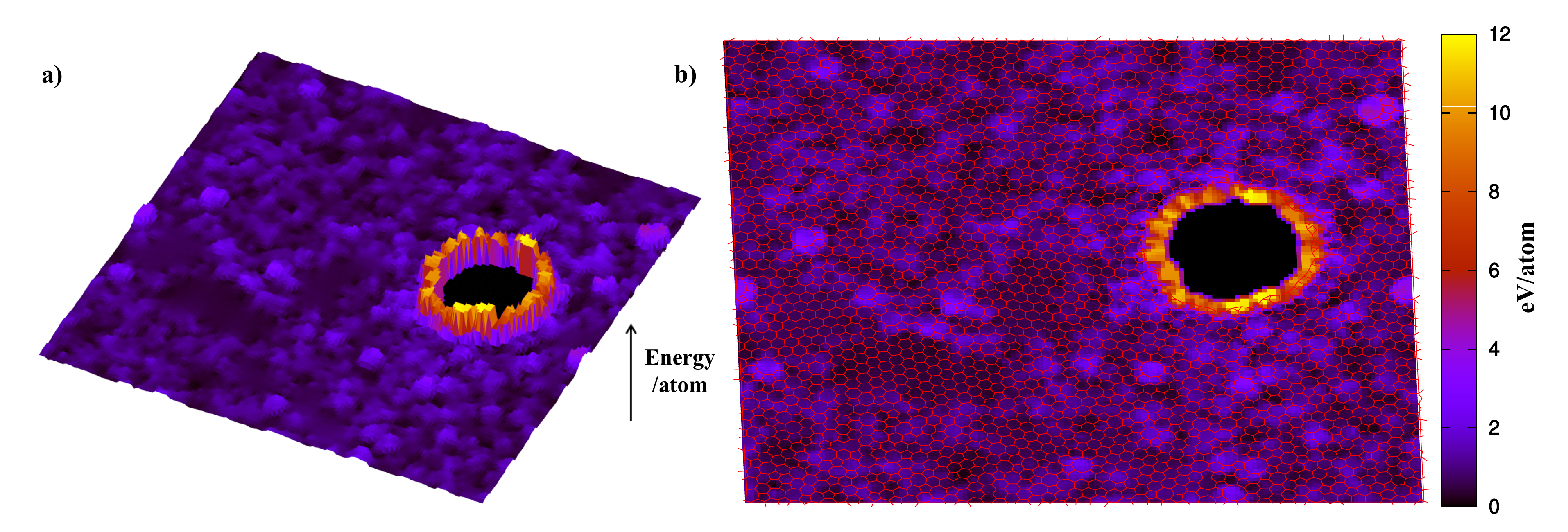}
\caption{Local energy distribution in the sample with a void radius of $16.92 $ \AA. In our samples, we define a local energy per atom (eV/atom) as follows: contributions due to two-body interactions are equally divided over the two interacting atoms, and contributions due to the three-body (angular) interactions are attributed to the central atom. Thus, the sum of the local energy over all atoms equals the total energy of the sample. (a) Most of the energy is localized at the periphery of the void in the sample. (b) Energy distribution along with the structure of the sample shows the amorphous nature of the sample (dark blue regime around the hexagonal rings).}
\label{Fig-4}
\end{figure*}

To further characterize the void, we calculate the local energy distribution along the sample as shown in Figure \ref{Fig-4}. In our samples, we define a local energy per atom (eV/atom) as follows: contributions
due to two-body interactions are equally divided over the two interacting atoms, and contributions due to the three-body (angular) interactions are attributed to the central atom. Thus, the sum of the local energy over all atoms equals the total energy of the sample. This definition of local energy helps us to visualize the local degree of mechanical and structural relaxation in the sample. The bulk of the total energy is concentrated along the periphery of void as shown in Figure \ref{Fig-4} b).  
\\
The summation of the local energy in the sample from the center of the void for $r \geq r_{v}$ can be given as
\begin{equation} \label{eq:Sum_energy}
\int^r_{r'=0} E(r') dr'=a\pi(r^{2}-r_{v}^{2}) + 2 \pi r_{v}\sigma(r_v).
\end{equation}

\begin{figure*}
\centering
\includegraphics[scale=0.44]{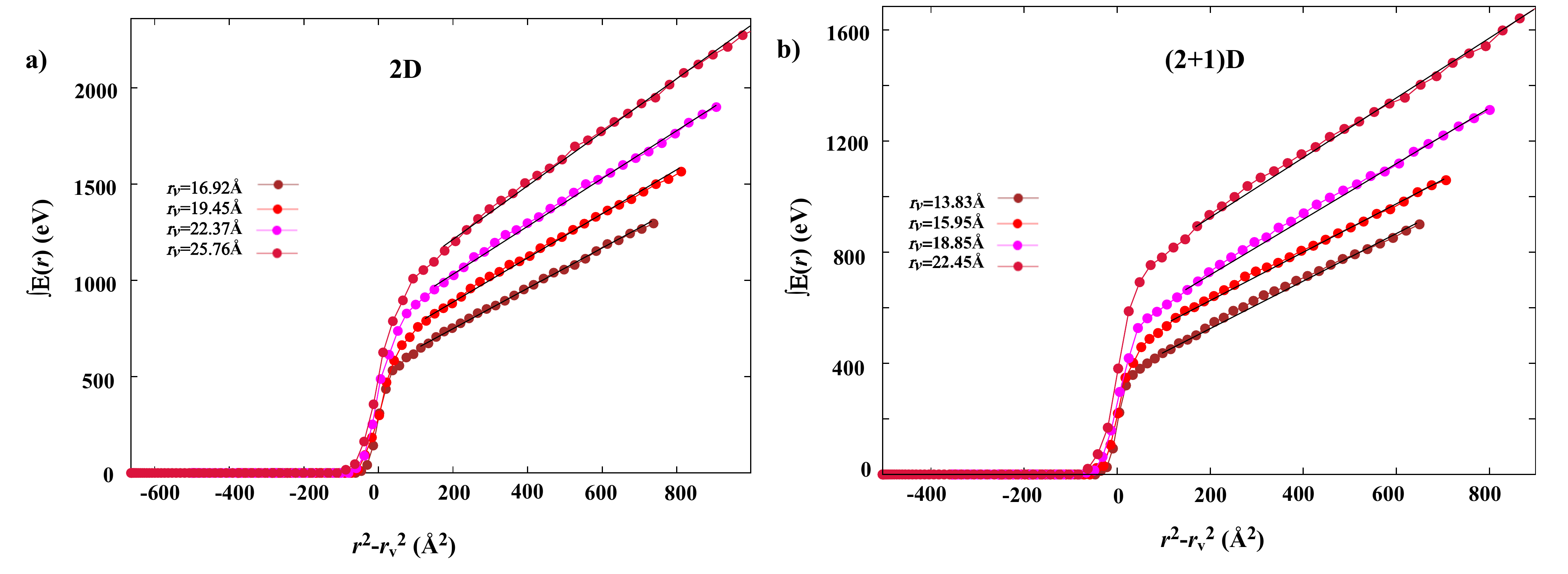}
\caption{The total summation of the elastic energy  $\left(\int E(r)\right)$ in the sample from the center of the void is plotted as a function of square of the distance $(r^2 - r_{v}^{2})$ for both flat and buckled samples. The summation of the energy in the sample from the center of the void can be written as ($r \geq r_{v}$) : $a \times \pi (r^2 - r_{v}^{2}) + 2 \pi r_{v} \sigma$; where $r$ is the distance from the center of the void, $r_{v}$ is the radius of the void, $\sigma$ is the line tension and $a$ is a fitting parameter. We fit these plots from $(r_{v}+2)$ to $(r_{v}+15)$ with a straight line to extract the values of $\sigma$ for different radius of the voids. Due to the additional relaxation the third direction in buckled samples the value of line tension of void is lower than that of the flat case.}
\label{Fig-5}
\end{figure*}

Here, $r$ is measured in \AA~ from the center of the void, $a$ is a fitting parameter in eV/\AA$^{2}$, and $\sigma(r_v)$ is the line tension (eV/\AA) along the void as a function of radius of the void $r_v$.
\\
We calculate the summation of the energy $\left(\int E(r)\right)$ as a function of $(r^{2}-r_{v}^{2})$ at different values of $r_v$ for both flat and buckled amorphous graphene samples as shown in Figure \ref{Fig-5} a) and b) respectively. By fitting a straight line with our numerical simulation data points in the plot, we extract the value of line tension $\sigma(r_v)$, from the range of ($r_{v}+2$) to ($r_{v}+15$) of $r$. We observe the finite size scaling in the line tension as a function of radius of the void $r_v$ for both flat and buckled samples. In general, the line tension ($\sigma(r_v)$) of a void can be written as
\begin{equation} \label{eq:line tension1}
\sigma(r_v)=\sigma(\infty)+f(r_v).
\end{equation}
Here $\sigma(\infty)$ is the value of the line tension in the thermodynamic limit and $f(r_v)$ is the nature of the finite size scaling in the line tension.

We plot the line tension as a function of inverse of the radius of the void and fit with a straight line for both flat and buckled cases as shown in Figure \ref{Fig-5}. The value of the $\sigma(r_v)$ in the thermodynamic limit approaches to $6.95$ eV/\AA~ and $6.61$ eV/\AA~ for flat and buckled amorphous graphene, respectively. To get a better understanding of the finite size scaling of the line tension we develop an analytical model which is discussed in the next section. \\

\begin{figure*}
\centering
\includegraphics[scale=0.44]{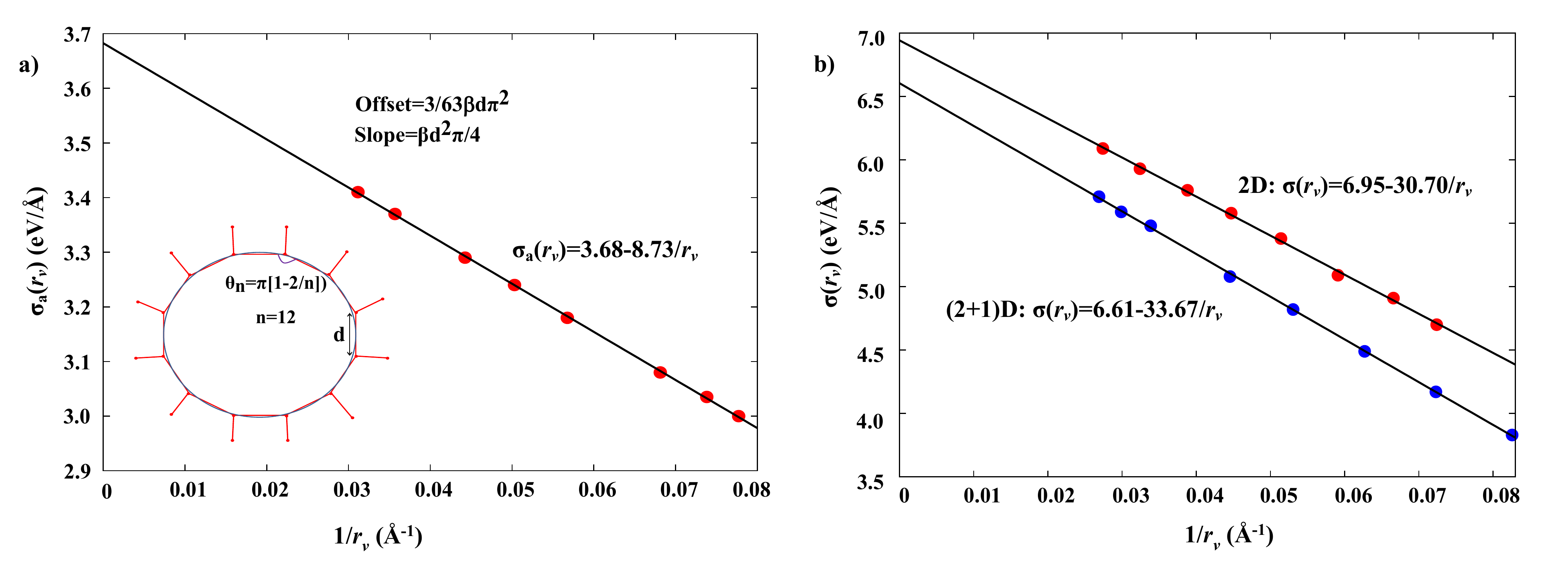}
\caption{The line tension of the void ($\sigma(r)$) is plotted as a function of inverse of the radius of the void and fitted with a straight line for both flat and buckled samples. (a) An analytical model based on three fold isotropic do-decagon with edge length of 1.42 \AA~ (crystalline carbon-carbon bond length in graphene) to numerically calculate the lower limit of the line tension is purely based on bond shearing. Analytical expression (Eq. \ref{eq:sigma_polygon}) suggests that there is $1/r_{v}$ finite size correction in the line tension which is consistent with our full simulations. (b) Finite size scaling behavior of line tension in both flat (red colored dots) and buckled (blue colored dotes) amorphous graphene. These data points at various values of radius of the voids are then fitted with a straight line (black line) to show that the line tension scales as inverse of the radius of the void and in the thermodynamic limit achieves the value of 6.95 eV/\AA~ for flat and 6.61 eV/\AA~ for buckled graphene samples.}
\label{Fig-6}
\end{figure*}

\textbf{Analytical model for the finite size scaling of the line tension}  
\\
To understand the energetics of a void in the amorphous sample, we study a simple analytical model of a polygon having $n$ equal length ($d$) edges as shown in Figure \ref{Fig-6} a). The angle between two consecutive edges can be written as 
\begin{equation} \label{eq:angle}
\theta_{n}=\pi \left[1-\frac{2}{n}\right] .
\end{equation}  

The effective energy of the polygon ($E_p$) calculated from Hamiltonian defined for graphene (Eq: \ref{eq:Potential}) will only be determined by three-body energy term (bond shearing therm) since the two body energy contribution will be zero (equal length edges) and can be written as
\begin{equation} \label{eq:energy_polygon}
E_{p}=\beta d^{2} \pi^{2} \left(\frac{3n}{63}-\frac{1}{2}\right) .
\end{equation}
\\
The area of a polygon is given by
\begin{equation} \label{eq:area_polygon}
A_{p}=\frac{nd^{2}}{4\tan(\pi/n)} .
\end{equation}
\\
The effective radius of the polygon for large values of $n$ can be written as, where $\tan(\theta) \sim \theta$, 
\begin{equation} \label{eq:radius_polygon}
r_{p}= \frac{dn}{2\pi} .
\end{equation} 
\\
The line tension along the void is obtained from the following equation
\begin{equation} \label{eq:line_polygon}
2\pi r_{p} \sigma_{p}(r_{p}) = E_{p},
\end{equation}
\\
which gives
\begin{equation} \label{eq:sigma_polygon}
\sigma_{p}(r_{p}) = \frac{3 \beta d \pi^{2}}{63}-\frac{\beta d^{2} \pi}{4r_{p}}.
\end{equation}
\\
This equation sets the lower limit of the line tension of a void in completely flat samples. In this equation we also observe the $1/r$ finite size scaling of the line tension of the void which is in the excellent agreement with our full simulations on amorphous graphene. In reasonable approximation, the lower limit of the line tension is set by the shear modulus of the sample. Since the line tension scales linearly with $\beta$, the energetics of the void also behaves linearly with the shear modulus of the sample. This analytical argument should be equally valid for other two-dimensional materials such as h-BN, phosphorene, borophene and other transition metal dichalcogenides. 

In our simulations we also tried to capture the effect of the bulk modulus on the energetic of the voids in two-dimensional materials. In particular we changed the values of $\alpha$ in the Hamiltonian (Eq: \ref{eq:Potential}) for graphene and minimized the energies of the various amorphous samples with voids. Counter-intuitively we observe that the two-body energy term increases by reducing the bulk modulus and vice versa. However the total energy of the sample (combined with stretching energy Eq: \ref{eq:Stretching}) decreases since reduction in bulk modulus makes sample less stiff and therefore more stretchable. These observations on the dependency of the energetic of the void in two-dimensional structures might be useful in strain engineering and tailored applications of these materials.\\   

\textbf{Critical radius of the void in 2D materials}
\\
We have shown that the line tension of the void scales as the inverse of the radius of the void and thus the expression can be written as 
\begin{equation} \label{eq:line tension}
\sigma(r_v)=\sigma(\infty)+\frac{C}{r_v}.
\end{equation}
Here $C$ is a fitting parameter which has a value of $-30.70$ \AA~ and $-33.67$ \AA~ for flat and buckled graphene, respectively, as shown in Figure \ref{Fig-5} b). Eq: \ref{eq:total_energy} can be rewritten as
\begin{equation} \label{eq:total_energy1}
F(r_v)= 2 \pi r_{v}\sigma(r_{\infty}) + 2 \pi C - \pi r_{v}^{2} P.
\end{equation}
\\
To calculate the critical radius of the void $r_{v(cri)}$ we put $\partial F(r_v)/ \partial r_v = 0$ and get
\begin{equation} \label{eq:Critical_radius}
r_{v(cri)}=\frac{\sigma_{\infty}}{P}.
\end{equation}
Hence, the total free energy at the critical radius of the void can be written as
\begin{equation} \label{eq:Max_total_energy}
F(r_{v(max)})= \frac{\pi \sigma_{\infty}^{2}}{P} + 2 \pi C.
\end{equation}
The values of $\sigma_{\infty}$ from our numerical simulations are $6.95$ eV/\AA~ and $6.61$ eV/\AA~ for flat and buckled graphene respectively. The critical radius of the void for flat and buckled amorphous samples are $3.48$\AA~ and $3.31$\AA~ respectively.\\

\section{CONCLUSIONS}
In conclusion, we show the effective mechanism and energetics of the rupturing process via void formation of amorphous (small domain polycrystalline) graphene under mechanical stretching. In amorphous samples even at 3-5\% of the stretching, samples start to rupture via formation of large voids whereas in crystalline graphene the stretching can be much higher (upto 20\%). Using the concepts of classical nucleation theory combined with simulations, we find the critical radius of the void is simply the ratio between line tension and the applied pressure on the samples.The values of the critical radius of the void for flat and buckled graphene are 3.48\AA ~ and 3.31\AA, respectively at 2 eV/\AA${^2}$ pressure. Line tension of the void has a finite size correction which scales as the inverse of the radius of the void for both flat and buckled samples. Line tension is directly proportional to the shear modulus of the sample and therefore future predictions regarding to the energetic of the voids in other two-dimensional materials can be achieved with the help of our results. Our results provide significant insight into the structural and energetics aspects of the voids in the polycrystalline samples and have direct implications on the tailoring of the properties of graphene via strain engineering \cite{Pereira2009a,Guine2012}.

\newpage

\newpage

\section{Acknowledgement}
We acknowledge the support by FOM-SHELL-CSER program (12CSER049). This work is part of the research program of the Foundation for Fundamental Research of Matter (FOM), which is part of the Netherlands Organization for Scientific Research (NWO).

\bibliography{Void}

\end{document}